%% file: main.tex
\documentclass[
    superscriptaddress,
    longbibliography,
    twocolumn,
    amsmath,
    amssymb
]{revtex4-2}

\usepackage{newtxtext}
\usepackage[utf8]{inputenc}
\usepackage[english]{babel}
\usepackage{mleftright}
\usepackage{graphicx}
\usepackage{mathtools}
\usepackage{booktabs}
\usepackage[table]{xcolor}
\usepackage[version=4]{mhchem}
\usepackage{fancyvrb}
\usepackage{listings}
\usepackage{natbib}
\usepackage{bibunits}
\usepackage{algorithm}
\usepackage{algpseudocode}
\usepackage{amsmath}
\usepackage{xspace}
\usepackage{hyperref}
\hypersetup{hidelinks}
\usepackage{acronym}
\usepackage{comment}
\usepackage{units}
\usepackage{varwidth}
\usepackage{stmaryrd}
\usepackage{bbm}
\usepackage{subfigure}
\usepackage{mathtools}
\usepackage{amssymb}
\usepackage{tabularx}
\usepackage{hyperref}
\usepackage{xcolor}
\usepackage{verbatim}

\makeatletter
\def\maketitle{
\@author@finish
\title@column\titleblock@produce
\suppressfloats[t]
}
\makeatother

\defaultbibliography{bib}
\defaultbibliographystyle{naturemag}

\algdef{SE}[VARIABLES]{Variables}{EndVariables}
   {\algorithmicvariables}
   {\algorithmicend\ \algorithmicvariables}
\algnewcommand{\algorithmicvariables}{\textbf{Global variables}}

\definecolor{backcolour}{rgb}{0.95,0.95,0.92}
\definecolor{commentgrey}{rgb}{0.4,0.4,0.4}
\definecolor{deepblue}{rgb}{0,0.6,1.0}
\definecolor{lightgrey}{rgb}{0.99,0.99,0.99}
\definecolor{deepred}{rgb}{0.0,0.6,1.0}
\definecolor{deepgreen}{rgb}{1.0,0.0,0.6}

\newcommand{\context}{\mathcal{C}}
\newcommand{\fullcontext}{\mathcal{D}}

\newcommand{\pK}{$\text{p}K$\xspace}

\newcommand{\silabel}{Supplementary Information\xspace}

\begin{document}

\title{Embracing assay heterogeneity with neural processes \\ for markedly improved bioactivity predictions}

\author{Lucian Chan}
\email{lucian.chan@astx.com}
\affiliation{Astex Pharmaceuticals, Cambridge, UK}

\author{Marcel Verdonk}
\affiliation{Astex Pharmaceuticals, Cambridge, UK}

\author{Carl Poelking}
\email{carl.poelking@astx.com}
\affiliation{Astex Pharmaceuticals, Cambridge, UK}

\begin{abstract}
Predicting the bioactivity of a ligand is one of the hardest and most important challenges in computer-aided drug discovery. Despite years of data collection and curation efforts by research organizations worldwide, bioactivity data remains sparse and heterogeneous, thus hampering efforts to build predictive models that are accurate, transferable and robust. The intrinsic variability of the experimental data is further compounded by data aggregation practices that neglect heterogeneity to overcome sparsity. Here we discuss the limitations of these practices and present a hierarchical meta-learning framework that exploits the information synergy across disparate assays by successfully accounting for assay heterogeneity. We show that the model achieves a drastic improvement in affinity prediction across diverse protein targets and assay types compared to conventional baselines. It can quickly adapt to new target contexts using very few observations, thus enabling large-scale virtual screening in early-phase drug discovery.
\end{abstract}
\maketitle
\begin{bibunit}

The primary aim of drug discovery is to design compounds that are safe for the patient and efficacious against the disease. During preclinical development, efficacy is usually equated with potency: A drug that is a potent inhibitor of a target protein is assumed to be effective against the associated disease. Hence building potency while maintaining a safe chemical profile is the key objective in early-phase drug discovery. Models that can predict the binding affinity as a measure of potency of a small molecule against a protein target are therefore the subject of intense research efforts. Accurate and transferable affinity models could help escalate the drug discovery process by identifying relevant candidate molecules across ultra-large chemical spaces~\cite{vsynthes_sadybekov_2022,ultralarge_lyu_2019,exploration_warr_2022}, such as Enamine REAL~\cite{enamine} or GalaXi~\cite{galaxi}. Large-scale, standardized and open-access data collection, curation and storage, such as provided by ChEMBL~\cite{mendez19} and the Protein Data Bank (PDB)~\cite{berman00}, are pivotal as they open up the possibility of data-driven, low-compute, deep-learning-based alternatives to experimental or physics-based computational screening techniques.

Among these data-driven techniques, we conventionally distinguish between {\it ligand-based}~\cite{deepdta_ozturk_2018,nguyen20_dta,Daga23,metadta_2022} and {\it structure-based} approaches~\cite{frustration_volkov_2022,mbp_yan_2023,jones_improved_2021,hacnet_2023} that typically produce models that are either {\it local} (i.e., specific to a certain protein system) or {\it global} in scope. Here a model is said to be {\it structure-based} if it makes use of the three-dimensional conformation of an experimental or predicted protein-ligand complex, instead of just the topological structure of the ligand and the protein identity or sequence. Despite significant interest from the community, existing structure-based models still fall significantly short of medicinal-chemistry requirements, with typical root-mean-square errors of around 1.2-1.3 \pK units~\cite{frustration_volkov_2022} when evaluated on the PDBBind core set~\cite{Wang04}. Ligand-based approaches, by contrast, benefit from a much larger data volume than structural datasets currently provide. They have shown adequate or even superior performance in identifying hits in a binary classification setting, but are susceptible to structural bias and memorization~\cite{memorization_wallach_2018}. Many attempts have been made to develop ligand-based regressors~\cite{deepdta_ozturk_2018,nguyen20_dta,Daga23}, with model performance varying widely depending on the data scope, locality, quality and volume. In practice, such ligand-based models are frequently used to filter out inactive compounds from large molecular libraries, followed by an evaluation with more accurate physics-based simulation methods, such as free-energy perturbation. These physics-based approaches can, however, suffer from significant systematic errors, generally require long simulation times, and are with current computing capabilities challenging to deploy on a scale beyond a few hundred to thousand compounds.

\begin{figure*}
    \includegraphics[width=\textwidth]{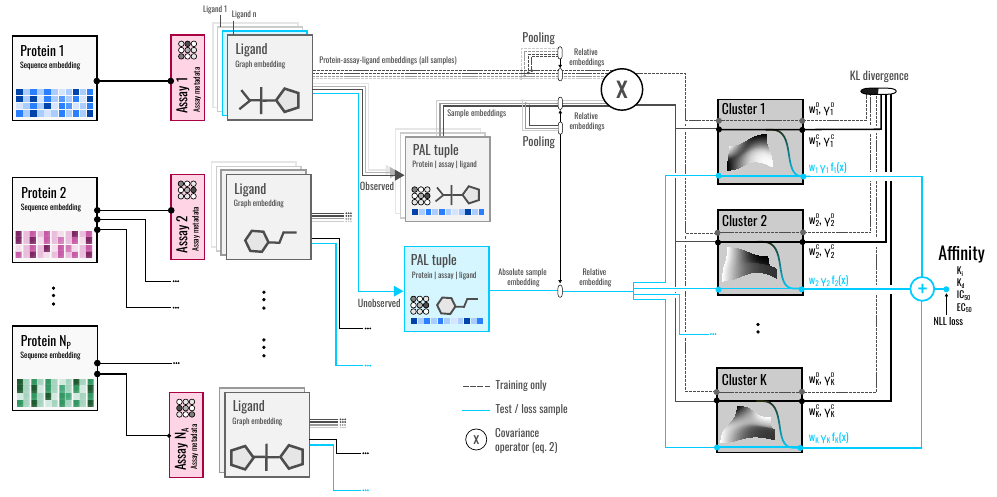}
    \caption{\textbf{Illustration of the MetaBind approach.} The model constructs an assay-specific SAR using a local support set of per-assay observations to predict affinities for unobserved protein-ligand pairs. In meta-learning terms, each assay is thus treated as a task, and assays are clustered into a small number of $K$ covariance groups (with, \text{e.g.,}, K=4). The subset of observed activities and the associated protein-ligand pairs are used to construct the SAR representation. At training time, this subset is randomly sampled from the available data on a per-assay basis, with the model trying to reconstruct the covariance structure associated with the full set of observations (dashed lines up top) from only a partial set sampled at random, as guided by a Kullback-Leibler loss term. The aggregation function, which is here based on a cross-covariance operator, as well as the protein and ligand embeddings are implemented using convolutional or (for ligands) graph-convolutional neural networks. The bioactivities of each protein-ligand pair and for each cluster are obtained via a multi-output neural network. Note that the assay-specific SAR is based, first and foremost, on the covariance structure inferred from the observed bioactivities of the context set. The assay metadata itself incorporates only the assay type (biochemical vs cell-based), in line with the notion that the formalism should be agnostic with respect to the source of the data heterogeneity, and that all condition variables are implicitly encoded into the bioactivities themselves.}
    \label{fig:architecture}
\end{figure*}

So why are data-driven models not performing better? The key issue is that the available bioactivity data are globally sparse and locally heterogeneous. This means that the models will struggle to identify relevant patterns across apparently {\it unrelated} assays (unrelated because they might target very different proteins); and, also, that they struggle to reconcile apparent inconsistencies among supposedly {\it related} assays (related because they might target similar proteins or assay similar compounds). For the latter, we use the term {\it heterogeneous} to describe the case where two seemingly equivalent assays measure systematically different structure-activity relationships (SARs). This heterogeneity -- or {\it between-assay} variability -- is therefore not caused by experimental uncertainty -- the {\it within-assay} variance. Instead, it is the result of differences in the experimental approach, assay type, assay conditions, and unobserved or undeclared hidden variables that affect the assay outcome: Consider, for example, cellular vs non-cellular and direct-binding vs functional assays; variations in pH, temperature, buffer composition or protein concentration; batch effects, solubility issues and non-specific binding; differences in the sensitivities and dynamic ranges of the experimental setups, or differences in the data analysis. There are thus a multitude of reasons why point-wise aggregation (i.e., the act of collating the results from different assays to create a single dataset per protein) introduces inconsistencies into the data, thus causing issues downstream. 

Previous analyses of mixed $\text{K}_{i}$ and $\text{IC}_{50}$~\cite{kramer12,kalliokoski13} data have attempted to estimate the experimental uncertainty from pairwise protein-ligand bioactivity relationships independently of the experimental approach. This line of analysis thus assumes that the variability is caused exclusively by uncertainty in the measurement itself, and does not separate the between-assay variability from the within-assay variance. As a rule of thumb, point-wise aggregation is meaningful only if the systematic variability between assays is far smaller than the statistical variability within. This implies that point-wise aggregation is particularly problematic when aggregating assays with different endpoints ($\text{IC}_{50}$ or $\text{K}_i$, for example), as this would require a conversion rule for the assay output~\cite{cheng73} that in turn relies on knowledge of either the bimolecular interaction kinetics or of any relative offsets -- neither of which are typically available.\\

Due to these sometimes hidden complexities of bioactivity data, we argue that the data aggregation process can be learnt implicitly by the predictive model itself, and that the model should be agnostic of the exact source of the heterogeneity. Here we show that this objective can be met with a hierarchical meta-learning model that incorporates assay heterogeneity while improving data efficiency. The model design reflects the fact that in a drug-discovery setting, we routinely find ourselves faced with a few-shot learning problem, as at best a few hundred to thousand molecules are made and tested during the typical life cycle of a discovery project. Importantly, the model is able to infer an assay-specific SAR from only a small number of observed data points in a few-shot manner, and is therefore able to `translate' results from one assay into the context of another. This assay-specific approach differs conceptually from previous meta-learning attempts and produces, as we show, a drastic improvement in predictive performance.

\section{Framework}
In this section we describe a meta-learning approach that models assay heterogeneity by exploiting the underlying correlation structure across diverse sets of assays. The resulting model, dubbed MetaBind, applies meta-learning at the assay level (each experiment thus constituting a separate task) to learn a local assay-specific SAR; it is a clustered multi-task neural process that can be viewed as an extension of standard neural processes~\cite{garnelo18b}.

Conventionally, to construct a neural process, task-specific samples, denoted as context data $\context$, are collected in order to infer the statistics of the associated domain. Based on these domain-wide statistics learned from the context, the model predicts the distribution $p(y|x, \context)$ of a target output $y$ for a given input $x$. In our case, the context data consists of the ligand structures and read-outs from a given assay. The learned statistics not only capture information on the underlying SAR, but also implicitly the experimental conditions. Using this framework, ligand affinities and uncertainties in the predictions can therefore be inferred independently for each assay, in contrast to existing models where the assay heterogeneity is neglected during training. The statistics generated from the context data is critical as it has to encapsulate the functional form of the SAR and be rich enough to identify variability across assays. Nevertheless, despite their flexibility, scalability and well calibrated uncertainty estimation, standard neural processes are designed to model data from only a \textit{single} stochastic process, which render them unsuitable for bioactivity data collected using various techniques (\text{e.g.,} biophysical, biochemical or cell-based assays), with various endpoints (\text{e.g.,} $\text{K}_{i}$, $\text{K}_{d}$, $\text{IC}_{50}$ and $\text{EC}_{50}$) and variable experimental conditions. 

A simple way of incorporating heterogeneity among related or equivalent assays would be to estimate an affinity offset from protein-ligand pairs that are shared across experiments (i.e., calibration or reference compounds). Such pairs are, however, few and far between within mixed-origin databases such as ChEMBL. Therefore, instead of estimating the offset directly (and thus limiting the model to constant-offset heterogeneities), the idea is to model the {\it relative} change within assays while incorporating information from related experiments in a mechanistically agnostic and functionally flexible manner. This is why here we introduce a clustered multi-task neural process (Fig.~\ref{fig:architecture}) that represents the bioactivity as a linear combination of a learnt reference affinity value $\tilde{y}^{\text{ref}}$, and $K$ \textit{independent} stochastic processes $f_{k}$ with scaling factor $\gamma$ that model the relative SAR within the assay. I.e., we use the ansatz
\begin{equation}\label{eqn_clusternp}
    \begin{split}
    y_{ia} = &\ \ y_{a}^{\text{ref}} + \sum_{k=1}^{K} w_{ka}\big(\gamma_{a} f_{k}(x_{ia} - x_{a}^{\text{ref}}) \big)  + \epsilon_{ia},\\
    \text{s.t.}  
    & \sum_{k=1}^{K}w_{ka} = 1,
    \end{split}
\end{equation}
where $x_{ia}$ is the representation of protein-ligand pair $i$, and $y_{ia}$ is the associated bioactivity relationship for assay $a$; $\epsilon_{ia}$ is an error term that follows a Gaussian distribution with zero mean and constant variance. The scales $\gamma_a$ and weights $w_{ka}$ are learnable functions (see {\it Methods} for details).

This formulation assumes that the assays can be clustered into a small number of covariance groups, and that the SAR within each group is defined by a mean function $f_{k}$ and assay-dependent scaling factor $\gamma_a$, as well as weight functions $(w_{1a},..., w_{ka})$ that quantify the relevance of each assay in the respective clusters. This functional structure makes the key difference to existing models, as assays are not aggregated in a point-wise fashion before the model is constructed, but instead aggregated implicitly on the fly in a manner learnt by the model itself.

Note that the cluster functions $f_k$ operate on a relative embedding $\tilde{x} = x_{ia} - x_a^{\text{ref}}$ to emphasise the variability within an assay relative to a reference calculated over the context $\context$,
\begin{align}
    (x_{a}^{\text{ref}}, y_{a}^{\text{ref}}) &= \left( \frac{\sum_{i}^{N_{a}^{\context}}x_{i}}{N_{a}^{\context}},  \frac{\sum_{i}^{N_{a}^{\context}}y_{ia}}{N_{a}^{\context}} \right).
\end{align}
To effectively aggregate functionally related assays, one could model the joint distribution of $(\tilde{x}, \tilde{y})$, \textit{i.e.,} $P_{\tilde{x}\tilde{y}}$, through mean embedding of concatenated samples, $\phi(\tilde{x}, \tilde{y})$. Here we instead represent this joint distribution using the cross-covariance operator $C_{XY}$\cite{gretton20} over the context $\context$:
\begin{equation}\label{eqn:cross_covariance}
    \begin{split}
    C_{XY}^a & = \frac{1}{||\context_a||} \sum_{i \in \context_a} \phi_{x}(\tilde{x}_{ia}) \otimes \phi_{y}(\tilde{y}_{ia}) \\
    & = \frac{1}{||\context_a||}\Phi_{x}^{T}\Phi_{y}^a \in \mathbb{R}^{h \times d},
    \end{split}
\end{equation}
where $\otimes$ denotes the outer product, $\phi_{\tilde{x}}$ and $\phi_{\tilde{y}}$ are feature maps for covariates and dependent variables, respectively; $h$ and $d$ are the output dimensions of $\phi_{\tilde{x}}$ and $\phi_{\tilde{y}}$. This cross-covariance is then flattened to parameterize the learned statistics via the weights $w_{ka} = w_{k}(C^a_{XY})$ and scale parameter $\gamma_{a} = \gamma(C^a_{XY})$. Conceptually, similarity of the SARs of two assays $a$ and $b$ implies similar statistics, $P_{\tilde{x}\tilde{y}}^a \simeq P_{\tilde{x} \tilde{y}^b}$, and thus similar functional parameters $w_{k}$ and $\gamma$.

This multi-modal approach with relative embeddings ensures that the neural process can capture non-trivial data heterogeneities. In the simple case that two assays differ by a mean shift caused by, for example, an annotation error that substituted nanomolar with micromolar units, the change is fully encapsulated in the reference affinity $y^{\text{ref}}$, whereas the induced shape parameters are unaffected, as the covariance structure $C_{XY}$ remains unchanged. Concept shifts in the SAR, by contrast, would trigger adjustments in $(\mathbf{w}, \gamma)$, in addition to potential changes in the reference itself.

In this study, we use bioactivity data from ChEMBL30 to train this neural process. Proteins and ligands are featurized via their sequence and molecular graphs, respectively. Please refer to the {\it Methods} section at the end as well as the Supplementary Information for further details on data processing, the featurization and encoding of protein-ligand pairs, the neural networks used to represent the shape parameters and cluster functions, the different loss terms or training protocol.

\section{Results and Discussion}

\begin{figure*}
    \includegraphics[scale=0.9]{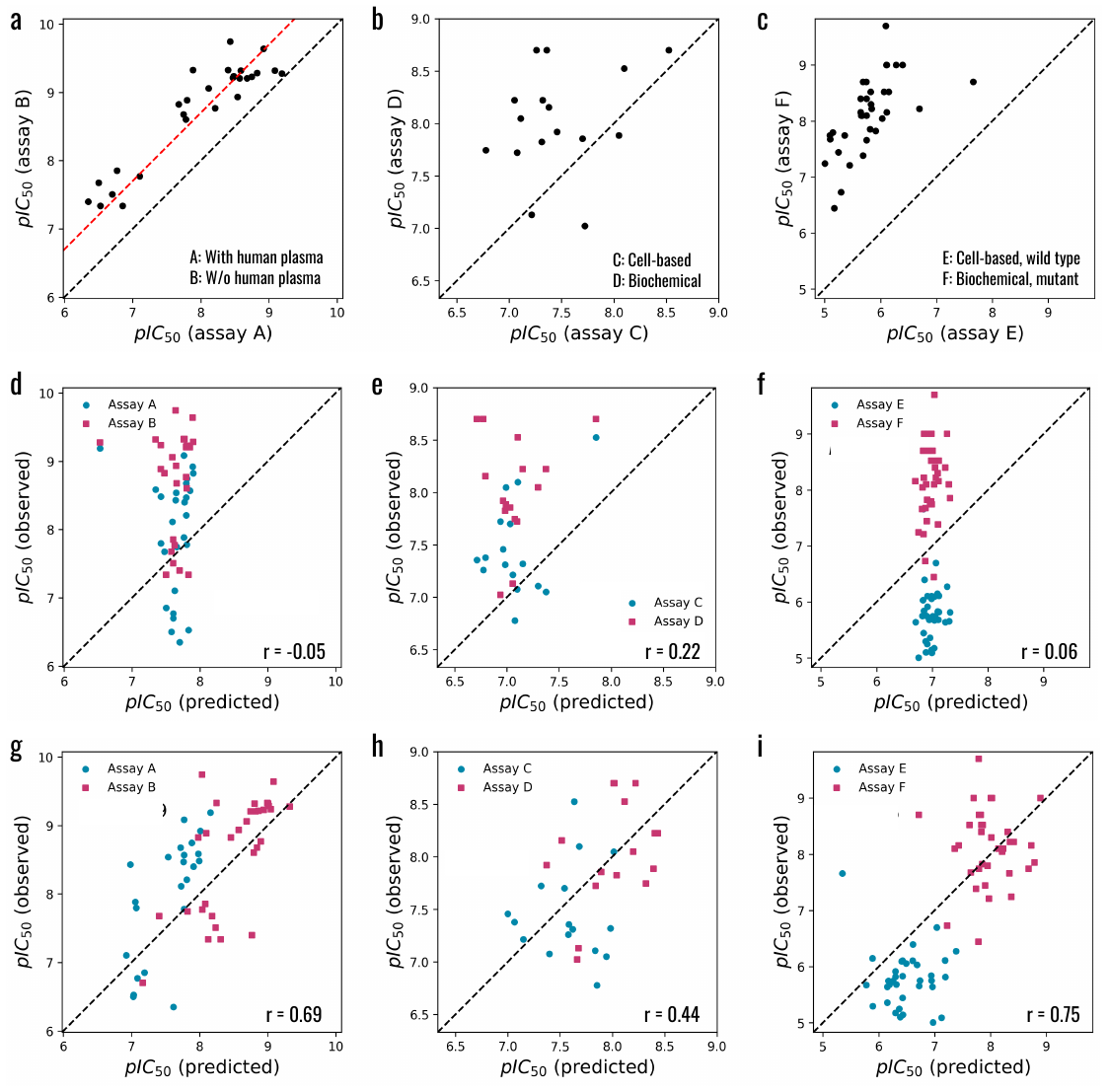}
    \caption{\textbf{Observed and predicted heterogeneities in structure-activity relationships.} (a-c) Correlation of the bioactivity read-out for congeneric series of three different assay pairs. The pairs (A,B) and (C,D) share the same protein variants, but differ in assay type and conditions. The assays of pair (E,F), on the other hand, differ in protein variant, with assay E investigating the wild type, and F a singly mutated variant thereof. The first assay pair (a) exhibits a strong correlation between the read-outs with, however, a noticeable systematic affinity offset. The second pair (b) does not show any significant correlation. The third pair (c) exhibits both an offset and rescaling of the bioactivities across the partners. (d-f) Baseline predictions for the series of assays A-F obtained with a standard graph-convolutional model. The baseline reproduces neither the activity trends within the assays, nor between the partners. (g-i) Affinity predictions obtained with MetaBind, demonstrating improved performance and at least partial reconstruction of the heterogeneity for all three pairs.}
    \label{fig:assay_hetero}
\end{figure*}

\begin{table*}

\begin{tabularx}{\textwidth}{X|XX|XX}
\toprule
\multicolumn{1}{c|}{} & \multicolumn{2}{c|}{\textbf{Paired (biochemical)}} & \multicolumn{2}{c}{\textbf{Chronological (biochemical)}} \\
\multicolumn{1}{c|}{} & \centering Task RMSE  & \centering Task MAE & \centering Task RMSE & \centering Task MAE\arraybackslash \\ \midrule
Baseline & \centering 1.125 (0.477) & \centering 0.981 (0.465)  & \centering 1.150 (0.477) & \centering 1.008 (0.480)\arraybackslash \\
Baseline (local) & \centering 1.102 (0.462)  & \centering 0.960 (0.450) & \centering 1.129 (0.464) & \centering 0.988 (0.466)\arraybackslash \\
MetaBind & \centering \textbf{0.794 (0.251)} & \centering \textbf{0.663 (0.231)}  & \centering \textbf{0.828 (0.270)} & \centering \textbf{0.689 (0.243)}\arraybackslash  \\ 
\end{tabularx}

\vspace{0.5cm}

\begin{tabularx}{\textwidth}{X|XX|XX}
\toprule
\multicolumn{1}{c|}{} & \multicolumn{2}{c|}{\textbf{Paired (cell-based)}} & \multicolumn{2}{c}{\textbf{Chronological (cell-based)}} \\
\multicolumn{1}{c|}{} & \centering Task RMSE  & \centering Task MAE & \centering Task RMSE & \centering Task MAE\arraybackslash \\
\midrule
Baseline  & \centering 1.160 (0.493) & \centering 1.023 (0.500) & \centering 1.171 (0.497) & \centering 1.030 (0.498)\arraybackslash  \\
Baseline (local)  & \centering 1.137 (0.484)  & \centering 1.001 (0.487)  & \centering 1.152 (0.485) & \centering 1.012 (0.487)\arraybackslash \\
MetaBind   & \centering \textbf{0.810 (0.236)} & \centering \textbf{0.672 (0.210)} & \centering \textbf{0.854 (0.282)} & \centering \textbf{0.712 (0.257)}\arraybackslash  \\
\end{tabularx}

\vspace{0.5cm}

\begin{tabularx}{\textwidth}{X|XX|XX}
\toprule
\multicolumn{1}{c|}{} & \multicolumn{2}{c|}{\textbf{Paired (all)}} & \multicolumn{2}{c}{\textbf{Chronological (all)}} \\
\multicolumn{1}{c|}{} & \centering Task RMSE  & \centering Task MAE & \centering Task RMSE & \centering Task MAE\arraybackslash \\
\midrule
Baseline   & \centering 1.110 (0.443) & \centering 0.974 (0.436) & \centering 1.131 (0.468) & \centering 0.992 (0.469)\arraybackslash \\
Baseline (local)  & \centering 1.090 (0.436)  & \centering 0.955 (0.430)  & \centering 1.113 (0.457)  & \centering 0.974 (0.459)\arraybackslash \\
MetaBind         & \centering \textbf{0.763 (0.234)} & \centering \textbf{0.632 (0.208)}   & \centering \textbf{0.803 (0.252)} & \centering \textbf{0.668 (0.227)}\arraybackslash   \\
\end{tabularx}
\caption{\textbf{MetaBind vs baseline performance in paired-assay and chronological test settings}. The baseline models are constructed using pre-aggregated training data. {\it Baseline} does not use any additional test data during predictions, as opposed to {\it Baseline (local)}, which performs a local update of the model given the assay context data. For each test setting (paired and chronological), we furthermore distinguish between biochemical and cell-based data partitions. MetaBind outperforms both baselines by a sizeable margin across all splits and test scenarios. Note the value in brackets, which indicates the standard deviation of the per-assay RMSEs and MAEs (not the error of the mean, which is far smaller). The MetaBind approach thus improves not only the predictions themselves, but also reduces (super-proportionally) the variability in performance across assays. Furthermore, it achieves the largest improvement in performance when aggregating biochemical and cell-based assays, despite the increased heterogeneity of the combined data.}
\label{tab:modelperformance}
\end{table*}

The MetaBind formalism proposed in this work is designed to handle assay heterogeneity in a functionally agnostic manner. This means that the approach does not make any specific assumptions how the SARs of two related assays differ. We illustrate this aspect first by showcasing how the model behaves for three examples that each reflect a different heterogeneity class. We will then proceed with a broader, statistical validation of the approach designed to indicate performance levels in a semi-prospective or prospective setting relative to conventional baselines. 

In qualitative terms, we can distinguish between three different grades of heterogeneity among related assays: Simple offsets, offsets combined with rescaling, and fundamental conceptual shifts in the SAR. Fig.~\ref{fig:assay_hetero} shows data for three different assay pairs that fall within one of these heterogeneity classes. Two assay pairs (Fig.~\ref{fig:assay_hetero}a-b) display distinct SARs despite the partners sharing the same protein-ligand series; the third pair (Fig.~\ref{fig:assay_hetero}c) compares experiments for the wild-type versus a mutated variant of the same target protein. The SARs for the first assay pair (panel a) are strongly correlated 
 with a Pearson correlation coefficient $r=0.93$, but a sizeable systematic offset of approximately 0.78 log units that might be the effect of human plasma that is present in one assay (assay A) but not the other (assay B).
 
 The second pair (panel b) compares a cell-based (C) and biochemical assay (D) for the same protein. The SARs are essentially uncorrelated with $r=0.26$, raising the question whether there is any meaningful information exchange possible between these assays. It is pairs like these that can cause severe issues during point-wise data aggregation, as the data samples are clearly drawn from two different distributions, and any agreement along the diagonal must be interpreted as coincidental. 
 
 The third pair (panel c) compares the SARs for two assays of different variants of the same protein: Assay E is a cell-based assay of wild-type B-Raf, whereas assay F is a biochemical assay of a singly mutated variant (V600E,  which is a key mutation in various cancer types). Although the measured bioactivities are clearly correlated with $r=0.63$, the SARs differ in terms of both scaling and offset. It is then impossible to determine without in-depth analysis what role the mutation plays in causing this difference, compared to the role of the variability in assay type, or the effect of a random heterogeneity due to unobserved or undeclared condition variables.

 So how do machine-learning models fare at predicting this variability in SAR? First, consider a standard graph-convolutional neural network (GCNN) that is trained on ChEMBL data preprocessed using point-wise aggregation rules. Expectedly, this simple baseline performs extremely poorly on this type of problem (see Fig.~\ref{fig:assay_hetero}d-f): Not only does the model ignore differences in SAR, but it also fails to produce any discernable correlation between its predictions and the experimental affinities {\it per assay}. Furthermore, its predictions tend to fall into a very narrow affinity range. Such behaviour can be caused by the learning dynamics of neural networks, which tend to produce smooth functions over the input space. As a result, small perturbations due to, e.g., minor changes in a lead series or single-point mutations of a protein, fail to produce a significant change in predicted SAR. This behaviour is aggravated further if the functional form of the predictor remains static, which is the case for conventional GCNNs.

\begin{figure*}
    \includegraphics[scale=0.95]{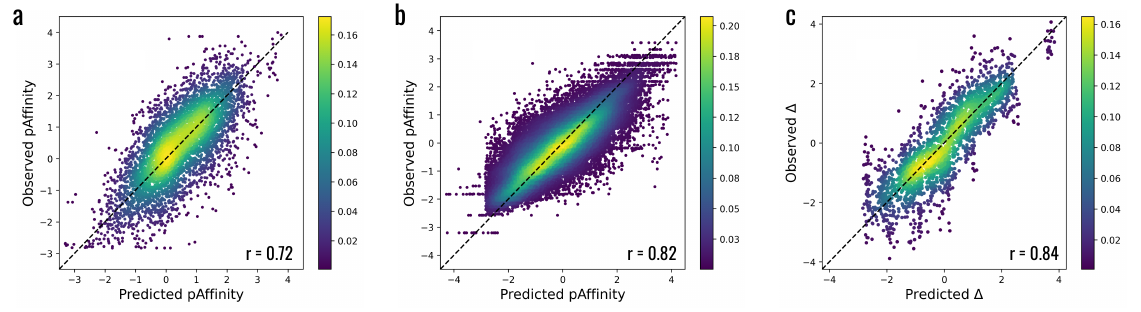}
    \caption{\textbf{Global correlation between predicted and observed bioactivities.} Tests based on the (a) paired-assay and (b) chronological split produce a strong Pearson correlation $r$ of $0.72$ and $0.82$ with the experimental affinities, respectively. The same holds for the (c) ligand heterogeneities extracted from the test set of the paired-assay split. These heterogeneities $\Delta$ (see their definition in eq.~\ref{eq:ligand_heterogeneity}) are not typically known, but here available by construction because of the congeneric series underlying the test set of the paired split. Note that panels a and b compare centred affinities to aid interpretation, with the same centring constant applied to all axes. }
    \label{fig:pred_obs_corr}
\end{figure*}

 The MetaBind model proves to be significantly more flexible in this regard (Fig.~\ref{fig:assay_hetero}g-i): Even though the intra-assay correlation of its predictions with the experimental observations is rather modest, the mean absolute errors for the different assays range between 0.41 and 0.85 \pK units. Importantly, the ligand heterogeneity, i.e., the difference in affinity for the same ligand $i$ but different assays $(a,b)$,
 \begin{align}
    \Delta_{ia,ib} = y_{ib} - y_{ia},
    \label{eq:ligand_heterogeneity}
 \end{align} 
 is reasonably well predicted with mean absolute errors (MAEs) of $0.35$, $0.48$ and $0.89$ \pK units for assay pairs (A,B), (C,D) and (E,F), respectively. These values should be compared to the mean absolute deviation of the measured ligand heterogeneities (which are observable here because of the joint ligand set) of $0.78$, $0.68$ and $2.33$. This suggests that the model is able to capture the heterogeneity at least partially. Analysis of the functional parameters (cluster weights $w$ and scales $\gamma$) gives further insight into the type of heterogeneity predicted by the model. E.g., for assay pair (A,B), the difference in predicted scaling is insignificant, $|\gamma_A - \gamma_{B}| \approx 0.005$, as is expected for a constant-offset heterogeneity. Meanwhile, the ranks of the cluster weights $w$ are identical within each pair, illustrating that the partners are predicted to fall within a similar SAR covariance class, despite perhaps what the experimental read-out from assay pair (C,D) suggests.

 These examples underline why `neural' data aggregation as performed by MetaBind is expedient. Still, the analysis so far was mostly anecdotal, and we will now pursue a broader validation of the approach. To this end, we again use ChEMBL30 data, this time to set up two different benchmark settings: A paired assay split and a chronological split. The former defines a test set of 190 assay pairs where the assays of each pair target the same protein, share at least 15 ligands, and have an intra-assay mean absolute deviation of at least $0.5$ \pK units. The latter uses a time split to divide the ChEMBL data into a training set (pre-2016) and a test set (post-2016), where the latter is made up of 178 unseen protein targets. Among the paired assays of the test set, only 34 exhibit a correlation $r > 0.7$, despite the Pearson correlation coefficient being invariant to constant offsets or rescaling of the SAR. For further details on the splitting procedure and dataset properties, see {\it Methods} and the \silabel. 

 The paired-assay split is designed to test the model's capability to infer the SAR heterogeneity, whereas the chronological split measures how well the model generalizes to new assay contexts. We include two baseline models as reference: First, a conventional GCNN -- denoted {\it Baseline} -- trained on pre-aggregated data, and a variant thereof that uses gradient updates on context data from the test set -- denoted {\it Baseline (local)}. The latter enables a fair comparison with MetaBind, as additional context data is used in an optimization-based meta-learning-style approach. Because MetaBind infers an assay-specific SAR from a context support set, a direct comparison with existing bioactivity and meta-learning models is not possible for conceptual reasons.

 The results of the comparison and MetaBind's test performance are summarised in Table~\ref{tab:modelperformance} and Fig.~\ref{fig:pred_obs_corr}a-b. We find that MetaBind outperforms the two baselines by a sizeable margin across all splits and metrics. The metrics we use are assay-level task RMSEs and MAEs (see {\it Methods} for their definition). For the test settings explored here, the mean task MAEs measured for the baselines range between $0.955$ to $1.171$, compared to $0.632$ and $0.854$ for MetaBind. For each of the two test scenarios (paired and chronological split) we additionally investigate different partitions of the ChEMBL data according to assay type (biochemical, cell-based, any): Cell-based assays capture the biological context more faithfully, but provide a less direct measure of ligand activity against a protein target, are less reproducible, and more susceptible to off-target effects. It is nevertheless reassuring that models built exclusively on cell-based data perform almost on par with those trained exclusively on biochemical data. Furthermore,  despite significant differences in chemical- and protein-space coverage of the biochemical and cell-based data partitions, MetaBind achieves a noticeable gain in performance when trained on a combined dataset. This is likely due to its enhanced ability to model the additional heterogeneity introduced into the data upon aggregation of diverse assay types. Overall, the good metrics and healthy correlation plots (Fig.~\ref{fig:pred_obs_corr}a-b) obtained with MetaBind indicate that the framework successfully constructs expressive SARs from local context data. The locally optimized baseline, by contrast, achieves only a very minor improvement over the standard GCNN.

 Finally, we stress that estimating offsets, scalings or other heterogeneities across two assays without duplicate compounds is typically an intractable problem. For the paired-test setting, however, we have the experimental ground truths for the heterogeneities available as by construction. This means that we can directly compare the predicted per-ligand shifts $\Delta_{ia,ib}$ to their experimental values. The correlation (Fig.~\ref{fig:pred_obs_corr}c) is excellent, with a Pearson $r$ of $0.84$ and a mean absolute deviation of $0.56$ \pK units. This is particularly encouraging as only five context observations were randomly selected from each assay to be supplied to the model, and not necessarily the same five within each pair. This indicates (as can be verified using dedicated tests, see the \silabel) that the covariance structure $C_{\tilde{X}\tilde{Y}}$ induced by the context set is robust. Still, if the signal-to-noise ratio is extremely low, the cluster assignment should be expected to become unstable, resulting in conservative predictions biased towards the sample mean. Potential solutions for this noisy regime could be to either increase the size of the context set, or provide additional metadata, for instance on binding sites or assay conditions.

\section{Conclusions}

Here we proposed a meta-learning formalism, MetaBind, that infers local, assay-specific SARs for protein-ligand affinity modelling by harnessing both local prior knowledge and global, diverse bioactivity data. The model enables a form of neural data aggregation to address the issue of assay heterogeneity: systematic differences in SAR produced by equivalent assays which, if unaccounted for, can have a severe and deleterious effect on model performance. We show that MetaBind produces exceptional results in different benchmark settings and in a few-shot manner, far outperforming conventional baseline models trained on pre-aggregated data, even when these are optimized for a specific assay context. MetaBind is thus able to adapt quickly to new target proteins, ligands, and assays, rendering it easily applicable to a drug-discovery setting.

Testing on paired assays highlights that the formalism is able to reconstruct in a robust way the heterogeneity even when presented with only minimal support data from each assay partner. Importantly, MetaBind does not assume a fixed functional form for the heterogeneity (such as constant offsets or rescalings), does not rely on knowledge of its causal origin, or on duplicate compounds shared across assays.

Data heterogeneity is generally considered a key issue that affects all major (bio-)chemical databases due to their mixed origin, diverse content and almost necessarily incomplete annotation. We therefore expect that formalisms such as MetaBind, which are based on an implicitly learnt, `smart' data aggregation function, will prove increasingly important in constructing transferable, performant models using community-driven, public-domain databases not just for bioactivity modelling, but also chemical reactivity, ADME-Tox predictions and beyond.

{\footnotesize

\section*{Methods}\label{main:method}

\subsubsection*{Framework}

\textbf{MetaBind architecture.}
We employed a clustered multitask neural process (CMTNP, eq.~\ref{eqn_clusternp}) to model the protein-ligand bioactivity because of its flexibility, scalability and capacity to produce well calibrated predictions. Derived from context data, the assay representation plays a crucial role in our framework as it provides the basis for how the model learns to aggregate data while constructing a local SAR. We use the cross-covariance operator in eq.~\ref{eqn:cross_covariance} to model the functional dependency of the SAR as probed by the different protein-ligand pairs of an assay, \textit{i.e.,} $P_{\tilde{X}\tilde{Y}}$. To infer appropriate scaling factors, the assay representation is normalised by the variance of the covariates. The full representation thus assumes the form
\begin{equation}
    x_a = \frac{\text{Cov}(\tilde{X}^a, \tilde{Y}^a)}{\text{Var}(\tilde{X}^a)},
\end{equation}
where $\text{Var}(\tilde{X}^a)$ is a diagonal matrix of variances ($\sigma^{2}_{\tilde{x}_{1}} \dots, \sigma^{2}_{\tilde{x}_{h}}$) estimated from context samples, with $h$ the latent dimension of the sample representation $x_{ia}$. Note that we also apply a global translation to the bioactivity values to ensure they are centered at zero.

Given the context data and the assay representation, the model jointly predicts shape parameters $(\alpha, \beta)$ and the cluster assignment $\mathbf{w}$ through a multi-output multilayer perceptron (MLP). This is followed by soft-plus and softmax activation functions over ($\alpha, \beta$) and $\textbf{w}$, respectively. Here we use $K=4$ covariance clusters. The scale parameter $\gamma = \frac{\alpha}{\beta}$ is calculated from the shape parameters of the associated gamma distribution $\text{Gamma}(\alpha, \beta)$; $w$ is modelled with a categorical distribution.

The model input, $x_{ia}$, captures information on the protein sequence and ligand structure of the protein-ligand pair $i$ within an assay $a$. Sequences are encoded using a convolutional neural network with a sliding window size of 20. Ligands are encoded using a graph-convolutional neural network operating on an atom-based molecular graph. The bioactivity predictions are formed by a multi-output MLP decoder, with $2K$ output channels, with the first $K$ channels corresponding to the mean function $f_{k}$, the second $K$ channels to the predicted variances $\sigma^{2}_{k}$. See the Supplementary Information for details on the input featurization and the network architectures of the encoders and decoder.\\

\textbf{Training protocol.} When training the model, we need to simulate an appropriate context set $\context$ using the available assay data. Chronological sampling would be ideal as this would most closely mimic the situation during hit-to-lead and lead optimization. Unfortunately, however, chronological rankings of the compounds are not usually available. Hence we resorted to on-the-fly sampling of random subsets of varying size to construct the context set, which trades in realism for improved data augmentation.

We train the model in an end-to-end fashion using the variational lower bound 
\begin{align}
    \text{log}\, p_{\theta}& (Y_{D}^{1:T}|X_{D}^{1:T}, C) \geq
    \mathbb{E}_{q_{\phi}} \text{log}\, p_{\theta} (Y_{D}^{t}| X^{t}_{D}, w^{t}, \gamma^{t}) \nonumber \\ &
    - D_{KL}(q_{\phi}(\gamma^{D}, w^{D}|D)||q_{\phi}(\gamma^{C}, w^{C}|C)) 
    \label{eqn:variationallowerbound}
\end{align}
where $D_{KL}(\cdot,\cdot)$ is the Kullback–Leibler (KL) divergence; $q$ and $p$ are the encoder and decoder networks, respectively. As in standard NPs, the KL divergence term encourages the network to predict scalings and weights from the context data, ($\gamma^{\context}, w^{\context}$) that are consistent with those from the full assay data ($\gamma^{\mathcal{D}}, w^{\mathcal{D}}$).\\

\textbf{Model implementation.}
The model was implemented in PyTorch~\cite{torch19} and optimized using ADAM~\cite{kingma17} with a learning rate of $1e^{-5}$. The graph-convolutional network was implemented using the Deep Graph Library (DGL)~\cite{dgl} package. RDKit~\cite{rdkit} was used to read, write and manipulate ligands. BioPython~\cite{biopython} was used to read, write and manipulate protein structures and sequences.\\

\textbf{Baseline models.}
In  contrast to traditional QSAR models, MetaBind learns to aggregate relevant assays instead of relying on hard-coded point-wise aggregation rules. Furthermore, its predictions are based on local support data (the context set) and hence, by design, assay-dependent. To fairly compare MetaBind with traditional QSAR approaches, we constructed two bioactivity models, {\it Baseline} and {\it Baseline (local)} with two distinct learning protocols. Like MetaBind, both models use convolutional and graph-convolutional neural networks to encode protein sequences and ligand structures, respectively. Furthermore, the encoder architectures are identical for both the baselines and the meta-learning model. The differences in architecture are therefore on the decoder side: Specifically, in the baseline models, the protein and ligand features are concatenated and fed into an MLP to predict the ligand bioactivity (see the Supplementary Information for details). 

Importantly, the baselines are trained using preprocessed training data. Duplicate observations are aggregated using their geometric mean if the deviation of the minimum and maximum observed values is less than 0.3 \pK units (corresponding to a threefold change in affinity) under the assumption that the endpoints are exchangeable; otherwise the observations are discarded. The two baseline models use identical network architectures, data pre-aggregation rules, and a mean-squared-error loss function for training, but differ in how they are applied to test data, as {\it Baseline (local)} uses a local training update over the context set to tune its weights to each test assay independently. This local version thus enables a fair comparison to MetaBind in that it produces an assay-specific SAR informed by local data. \\

\subsubsection*{Related Work}
\textbf{Meta-learning.} 
Meta-learning is an umbrella term for machine-learning approaches that aim to help a model adapt to new tasks using information on previous tasks. Various studies have applied meta-learning principles to molecular property prediction~\cite{han17,nguyen20,pappu20} and molecular optimisation~\cite{wang21} in low-data regimes. These meta-learning techniques can be broadly classified into model-based, optimisation-based~\cite{finn17} and metric-based~\cite{snell17} approaches. Model-based techniques aggregate information from previous tasks to extrapolate to new tasks and contexts. Optimisation-based approaches specialise a meta-learner on a new task using, e.g., gradient-based optimization on a support set. Metric-based approaches construct a similarity metric across tasks and use embedding-based queries to contextualize a support set.\\

\textbf{Multi-task learning.}
Multi-task models have a long-standing tradition in molecular-property prediction, and have also been developed for bioactivity modelling~\cite{martin19, whitehead19, pentina22}. These models generally treat each assay as a task and model all tasks jointly. This formulation avoids the assay heterogeneity effect introduced in point-wise aggregation, and explicitly exploits the correlation between assays to predict the bioactivity of new protein-ligand pairs. These approaches, however, typically require a moderate number of observations per task ($>50$) and some degree of compound overlap among the different assays, which renders them impractical to apply to low-data protein targets.\\

\textbf{Neural processes.} 
Neural Processes (NPs)~\cite{garnelo18b} learn a stochastic process over functions as solution to a supervised learning problem. They thus combine the adaptability of neural networks with the characteristics of Gaussian Processes. The neural process is constructed in two stages. First, it learns the statistics of a generic domain from a large sample set without committing to a specific learning task. Second, using these domain-wide statistics, it constructs a distribution over functions for a specific task from a small support set. An NP variant also includes a global latent variable, $z$, to account for the uncertainty in the predictions of the output for given input pairs ($x$, $y$)~\cite{kim2018}. Its application to molecular-property prediction has been explored recently~\cite{lee22,miguel22}. We note that standard NPs only model data from a single stochastic process and are thus designed to infer each task independently. There have been various efforts to extend NPs to a multi-task setting~\cite{kim2022}.\\

\subsubsection*{Data}

\textbf{Data filters.} ChEMBL 30~\cite{mendez19} was used throughout the study. We considered binding assays that are assigned to a single protein target with a ChEMBL confidence score of at least 9 (the maximum score). We included four endpoints: $\text{K}_{i}$, $\text{K}_{d}$, $\text{IC}_{50}$ and $\text{EC}_{50}$. The target species is {\it homo sapiens}. Here we do not consider censored activity data, i.e., we remove data samples with inequality labels $<$ or $>$. Moreover, we exclude assays that report identical activity values for more than five compounds, or those with an activity range of less than $0.5$ \pK units. Among the remaining set, we include only assays with more than ten exact measurements, and consider only ligands with heavy-atom counts $\leq 50$ and the following heavy-atom types: C, N, O, F, P, S, Cl, Br, I. The resulting datasets consists of 6968 assays, with a total of 150,159 measured affinities across 100,279 unique ligands and 830 distinct target proteins. Note that we only aggregate duplicates {\it within} assays, but keep duplicates across assays separate. See the Supplementary Information for a more detailed data summary.\\

\textbf{Metrics and data splits.} We evaluate the model performance at the task (i.e., assay) level using a task mean squared error (T-RMSE) and mean absolute error (T-MAE):
\begin{align}\label{task_level_metrics}
        \text{T-RMSE}_{a} &= \sqrt{\frac{1}{N_{a}}\sum_{i=1}^{N_{a}}(\hat{y_{ia}} - y_{ia})^{2}},\\
        \text{T-MAE}_{a} &= \frac{1}{N_{a}}\sum_{i=1}^{N_{a}}|\hat{y}_{ia} - y_{ia}|,
\end{align}
with task $a \in \{1, ..., n\}$, observations $i \in \{1, ..., N_{a}\}$, and $N_{a}$ the number of observations for task $a$; $\hat{y}$ is the predicted binding affinity (or bioactivity), $y$ is the observed binding affinity (or bioactivity); $|\cdot|$ indicates the absolute value. We additionally use the Pearson's correlation coefficient $r$ to assess the (global) correlation between the predicted and observed affinities.

To assess the degree to which a model is able to capture assay heterogeneity, we constructed a test set of assays pairs where the partners share identical target proteins together with a series of at least fifteen ligands. Additionally, we ensured that the mean absolute deviation of the bioactivities within the individual assays is $0.5$ log units or larger. We thus identified 100 assay pairs formed from among 190 unique assays as test set, and denoted the corresponding split as the {\it paired-assay} split. The affinities of the assay partners are not generally strongly correlated, see the Supplementary Information for details.

To assess how the models adapt to new assays and targets, we split the assays chronologically, with assays prior to 2016 entering the training set, assays of 2016 or later the test set. This results in a test set of 178 unseen target proteins. The median ligand similarity between training and testing set is 0.32. See further details in the Supplementary Information.\\

}

{\footnotesize

\textbf{Data availability.}
The datasets used in this study are available at \url{https://github.com/lucianlschan/metabind}\\

\textbf{Code availability.}
The source code and pre-trained models can be accessed at \url{https://github.com/lucianlschan/metabind}.\\

\textbf{Acknowledgements.} LC acknowledges funding from Astex through the
Sustaining Innovation Postdoctoral Programme. We thank Chris Murray, Davide Branduardi, Lisa Ronan, Tugce Oruc, Rudolfs Petrovs and Caroline Richardson for thoughtful comments on the manuscript. \\

\textbf{Author contributions.}
LC conceived the project. LC and CP designed the model. LC implemented the model, ran the experiments and performed the data analysis. LC, MV and CP devised the analysis and validation. LC and CP wrote the manuscript. CP and MV supervised the research. All authors reviewed the manuscript.\\

\textbf{Competing interests.}
The authors declare no competing interests.\\

}


\end{bibunit}
\newpage
\clearpage
\input{si}
\end{document}

%% file: si.tex
\title{SUPPLEMENTARY INFORMATION \\ \vspace{0.5cm} {Embracing assay heterogeneity with neural processes \\ for markedly improved bioactivity predictions}}
\maketitle

\setcounter{figure}{0}
\renewcommand{\thefigure}{S\arabic{figure}}
\setcounter{section}{0}
\renewcommand{\thesection}{S\arabic{section}}
\setcounter{subsection}{0}
\renewcommand{\thesubsection}{\Alph{subsection}}

\begin{bibunit}
\subsection{Data}~\label{si:chemblsummary}
\begin{figure*}
        \includegraphics[scale=1.0]{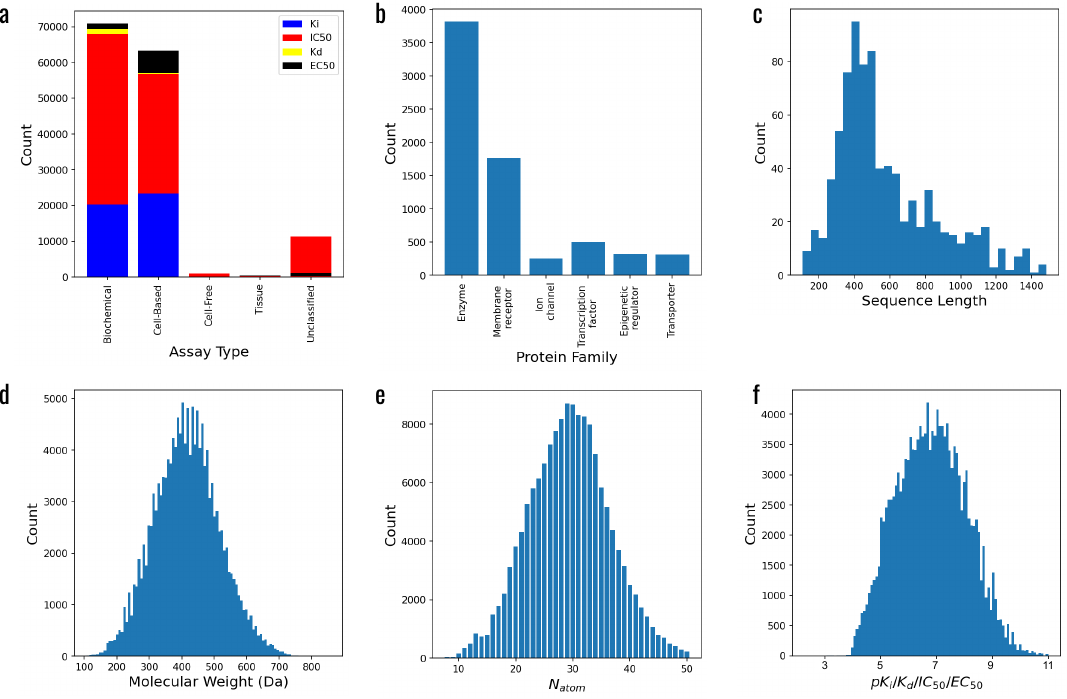}
        \caption{\textbf{Dataset summary.} 
        (a) Composition by assay type, with biochemical and cell-based assays being the predominant types. Half maximal inhibitory concentration ($\text{IC}_{50}$) and inhibition constant ($\text{K}_{i}$) together make up more than 90\% of the data volume.
        (b) Composition by protein target family. Enzymes are the predominant target class, followed by membrane receptors.
        (c) Composition by protein-sequence length. The median sequence length is 491.
        (d) Distribution of ligand molecular weights, ranging from \unit{107}{Da} to \unit{857}{Da}, with a median of \unit[415]{Da}.
        (e) Distribution of per-ligand heavy-atom counts. The median is 29.
        (f) Distribution of log-affinity values. The median is $6.8$.        
        }
    \label{sifig:chembl2dsummary}
\end{figure*}

Summary statistics for the assays selected from ChEMBL 30 for this study are provided in Fig.~\ref{sifig:chembl2dsummary}. The dataset includes 6968 assays, covering 830 target proteins with measured bioactivities for 100,279 molecules.  The molecular weight ranges from \unit[107]{Da} to \unit[857]{Da}, with a median of \unit[415]{Da}. The heavy-atom counts per ligand range from 8 to a maximum of 50 (as by construction). Enzymes are the predominant protein target class, followed by membrane receptors, as based on ChEMBL's classification schema. The median target sequence length is 471. The half maximal inhibition concentration, $\text{IC}_{50}$, and inhibition constant, $\text{K}_{i}$, are the most common endpoints, accounting for ca 63\% and 31\% of the data, respectively. The p$K$ values range from $-6.7$ (millimolar, mM) to $2$ (picomolar, pM) across all assays, with an average of $-2.21$.

\subsection{Notation and definitions} 
We denote by $\context^{a}$ and $\fullcontext^{a}$, with $a = 1,...,K$ , the context and the target data from assay $a$, respectively, where $K$ is the number of assays in the dataset. Further, we define $\context = \cup_{a=1}^{K}\context^{a}$ and $\fullcontext=\cup_{a=1}^{K}\fullcontext^{a}$. 

The neural network components used for the convolutional and graph-convolutional neural networks include activation functions, layer normalisation, convolution and pooling operations as defined below. \\

\subsubsection*{Activation functions} 
\noindent   Rectifying Linear Unit (ReLU)
        \begin{align*}
            \text{ReLU}(x) &= \begin{cases}
                          0 & \text{if $x\leq0$} \\
                          x & \text{if $x>0$} 
                          \end{cases}
        \end{align*}
    Softplus ($\beta$ = 1)
        \begin{align*}
            \text{Softplus}(x) &= \frac{1}{\beta}\log\big(1+\exp(\beta x)\big) 
        \end{align*}
    Tanh 
        \begin{align*}
            \text{Tanh}(x) &= \frac{\exp(x) -\exp(-x)}{\exp(x) + \exp(-x)} 
        \end{align*}
    Softmax 
        \begin{align*}
            \text{Softmax}(x) &= \frac{\exp(x_{i})}{\sum_{i=1}^{N}\exp(x_{i})} 
        \end{align*}

\subsubsection*{Convolutions}

\noindent Affine transformation
    \begin{align*}
        \text{Linear}(x) = Wx +b
    \end{align*}

\noindent Molecular GCNN layer
    \begin{align*}
        h_{i}^{l+1} = \sigma \left( b^{l} + \sum_{j\in N(i)}\frac{1}{c_{ij}}h^{l-1}W^{l} \right)
    \end{align*}
Here, $N(i)$ is the set of neighbors of node $i$, $c_{ij}$ is the square root of the product of the node degrees, $\sigma$ is an activation function, $l$ is the layer index, $W^{l}$ and $b^{l}$ are learnable weights and biases, respectively.

\subsubsection*{Pooling operators}
\noindent Max Pooling
    \begin{align*}
        \text{maxpool}(x^{i}) = \max_{k=1}^{N_{i}}\, (x_{k}^{i})
    \end{align*}
Sum Pooling
    \begin{align*}
        \text{sumpool}(x^{i}) = \sum_{k=1}^{N_{i}} (x_{k}^{i})
    \end{align*} 

\subsection{Architectures}~\label{SISec:Model}
\textbf{Ligand input features.} Atoms (nodes) and bonds (edges) of the molecular graph are featurized using a set of chemical properties as summarized in Tab. S3 and S4. Categorical features are encoded in a one-hot format, resulting in an overall input dimension of 36 for the atom and 6 for the bond feature vectors. RDKit was used to compute all atom and bond features.\\

\begin{table}
\centering
    \begin{tabularx}{\columnwidth}{l X l}
        \toprule
        \textbf{Feature} & \textbf{Description} & \textbf{Dimension} \\
        \midrule
         Atom type & Element (one-hot) & 12  \\
         Atom degree & Neighbour count (one-hot) & 7 \\
         Radical electrons & Radical-electron count & 1\\
         Formal charge  & Charge in $e$ & 1 \\
         Hybridization & $s$, $sp^{1}$-$sp^{3}$, $sp^{3}d$, $sp^{3}d^{2}$, other  & 7 \\
         Aromaticity & Aromatic indicator & 1 \\
         Hydrogen count & Explicit + implicit & 5 \\
         Chirality & CIP R and S indicators & 2\\
         \bottomrule
    \end{tabularx}
    \caption{\textbf{Atomic input features.} Atom node attributes used by all 2D molecular embedding networks.}
    \label{tab:atomfeature}
\end{table}

\begin{table}
    \centering
    \begin{tabularx}{\columnwidth}{l X l}
        \toprule
        \textbf{Feature} & \textbf{Description} & \textbf{Dimension} \\
        \midrule
        Bond type & Single -- triple, aromatic, conjugated & 5 \\
        Ring bond & Ring indicator & 1 \\
         \bottomrule
    \end{tabularx}
    \caption{\textbf{Bond input features.} Covalent edge attributes used by all 2D molecular embedding networks.}
    \label{tab:bondfeature}
\end{table}

\textbf{Protein input features.} The residues of the protein sequence are encoded in a one-hot fashion using the amino-acid type (AA), the sidechain topology and its physicochemical properties. The four side-chain topologies are: (i) no sidechain, (ii) linear side-chain, (iii) branched side-chain, and (iv) cyclic side-chain. The physicochemical properties capture the electrostatic and hydrophobic characteristics of the different AAs: positively charge (H, K, R), negatively charged (D, E); polar (N, Q, S, T); aromatic (F, W, Y); hydrophobic (A, I, L, M, V).\\

\begin{table}
    \centering
    \begin{tabularx}{\columnwidth}{l X l}
        \toprule
        \textbf{Feature} & \textbf{Description} & \textbf{Dimension} \\
        \midrule
        Amino Acid & 20 AA and 5 special characters & 25 \\
        Side Chain Topology & AA side-chain topology & 4 \\
        Properties & AA physicochemical properties & 6\\
         \bottomrule
    \end{tabularx}
    \caption{\textbf{Protein input features.} Amino-acid (AA) features used as input for the 1D protein-sequence embedding networks.}
    \label{tab:sequencefeature}
\end{table}

\textbf{Ligand representation.}
The ligand is represented as a graph, featurized as described above, and encoded using a 3-layer graph-convolutional neural network, with residual connections between each layer. To generate the ligand features, we apply max and sum pooling over the atom encodings, and concatenate the respective outputs to obtain the ligand embedding.\\

\textbf{Protein representation.}
The protein is represented as a sequence, featurized as described above, and encoded using a discrete cosine transform applied to the feature matrix, followed by a 1D convolution operation with kernel size 10 and stride 3. Note that we set a maximum sequence length of 1500 residues.\\

\textbf{Protein-ligand representation.}
The joint protein-ligand embedding vector is generated from the individual ligand and protein embeddings via concatenation and subsequent linear transformation.

\begin{table}
\begin{tabularx}{\columnwidth}{lXXX}
\toprule
\textbf{Model}               & \textbf{Input}                     & \textbf{Architecture} & \textbf{Dimensions}   \\ \midrule
Ligand Encoder      & L     & GCN          & (34,128,128)     \\ 
Protein Encoder     & P     & CN           & (32 ,32, 1)      \\ 
P-L Encoder         & P, L  & Linear       & (419,128)        \\ 
Cluster Decoder     & P-L,y & MLP          & (128,128,6)      \\ 
Bioactivity Decoder & P-L   & MLP          & (128,128,$h$)
\\
\bottomrule
\end{tabularx}
\caption{\textbf{Model architectures.} L, P, P-L, y denote the ligand, protein, protein-ligand and bioactivity, respectively. GCN, CN, Linear and MLP are the graph-convolutional and convolutional network, linear map and multi-layer perceptron, respectively. The values in brackets indicate the input, intermediate and output dimension, with (see bottom row) $h=8$ for MetaBind, $h=1$ for the baseline models. }
\end{table}

\subsection{Data splits}

\begin{figure*}
    \includegraphics[scale=0.95]{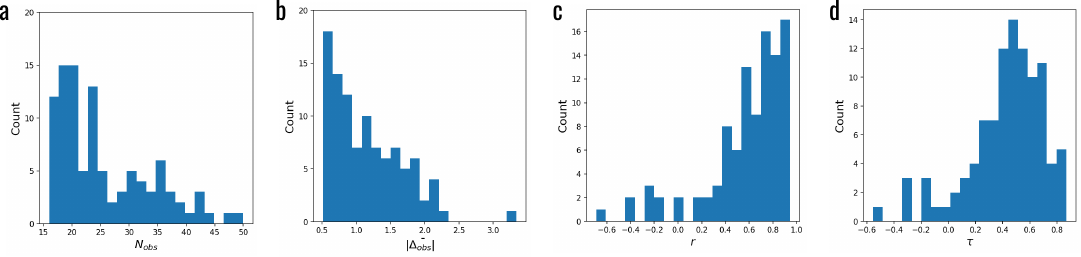}
    \caption{\textbf{Paired-assay test set.} (a) Distribution of the size of the shared ligand set (number of observations $N_\text{obs}$ per assay pair. (b) Distribution of the mean absolute deviation $\Delta$ measured for the assay partners. (c) Distribution of the Pearson correlation coefficient and (d) Kendal tau of the pair read-outs. Note that some assays pairs show a negative or statistically negligible correlation.}
    \label{sifig:paired_assay_split_summary}
\end{figure*}

\textbf{Paired-assay split.} 
The paired-assay test set contains 100 assay pairs generated from 190 assays. The distribution of the number of observations per assay pair is shown in Fig.~\ref{sifig:paired_assay_split_summary}a. The distribution of the mean absolute deviation within a pair (i.e., the mean of the absolute ligand hererogeneity $\Delta_{ia,ib}$) is shown in Fig.~\ref{sifig:paired_assay_split_summary}b, and has a median of $1.05$ log units. The majority of the assay pairs are positively correlated, as shown in Fig.~\ref{sifig:paired_assay_split_summary}c-d using the Pearson correlation $r$ and Kendall rank correlation coefficient $\tau$. There are multiple assay pairs, however, for which the correlation is negligible or even negative, see the tails of the distributions over $r$ and $\tau$.

\begin{figure*}
    \includegraphics[scale=0.8]{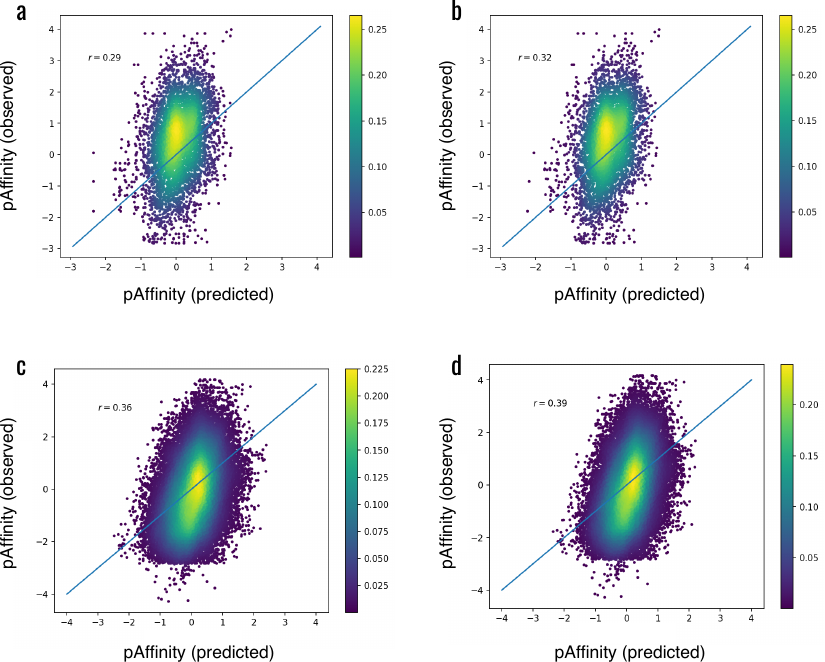} 
    \caption{\textbf{Baseline model performance.} Correlation plots comparing observed and predicted affinity values for the (a) standard baseline and (b) local baseline under the paired-assay split; (c) the standard baseline and (d) local baseline under the chronological split. To aid interpretation, all panels compare centred affinities, with the same centring constant applied to all axes. }
    \label{sifig:baseline_scatter}
\end{figure*}

The correlation between observed and predicted values achieved by the standard and local baseline models on the paired-assay test set is shown in Fig.~\ref{sifig:baseline_scatter}a and b, respectively. The correlation is weak, with a Pearson $r$ of $0.29$ and $0.32$, respectively. The local update on context data thus improved the model performance only slightly, limited presumably by the Lipschitz continuity/constant inherent to the neural network.\\

\begin{figure*}
\includegraphics[scale=0.8]{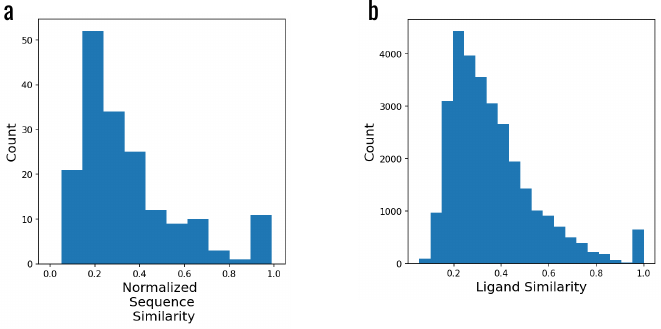}
\caption{\textbf{Chronological test set.} (a) Distribution of the maximum sequence similarity between unobserved protein targets in the test set with protein targets of the training set. (b) Distribution of the maximum ligand similarity between ligands of the training set and ligands of the test set for all observed protein targets. The median maximum similarity is $0.32$. Note that we used Tanimoto similarity over 2048-bit Morgan fingerprints of radius 2 to measure the ligand-ligand similarity.}
\label{sifig:chronological_split_summary}
\end{figure*}

\textbf{Chronological split.}
The test set of the chronological split contains 539 proteins, with 178 unseen targets. The distribution of the normalised sequence similarity between the unseen target proteins and observed target proteins is shown in Fig.~\ref{sifig:chronological_split_summary}a. Here the normalised sequence similarity is defined as the product of the identity score calculated from BLAST and the min-max ratio $\rho$ of the sequence lengths,
\begin{align}
    \rho_{AB} = \frac{\min(N^{A}_{AA}, N^{B}_{AA})}{\max(N^{A}_{AA}, N^{B}_{AA})}, \nonumber
\end{align} 
where $N^{A}_{AA}, N^{B}_{AA}$ are the lengths of sequences A and B, respectively. We used default parameters for the BLAST. For assays with targets already seen during training, we calculated the maximum ligand similarity between the training set and test set of those targets, with the resulting distribution of similarities shown in Fig.~\ref{sifig:chronological_split_summary}b. Even though some assays share identical ligands, most of the ligand sets differ considerably, as indicated by a relatively low median ligand similarity of $0.32$.

As for the paired-assay test set, both the standard and local baseline give poor fits with a limited dynamic range and Pearson $r$ of $0.36$ and $0.39$, respectively (see Fig.~\ref{sifig:baseline_scatter}c-d).

\begin{table*}[bp]
\begin{tabularx}{\textwidth}{clX}
\toprule
\textbf{Assay} & \textbf{ChEMBL ID} & \textbf{Description} \\ 
\midrule
\hspace{0.5cm} A \hspace{0.5cm}     & CHEMBL1039372 \hspace{1cm} & \begin{tabular}[c]{@{}l@{}}Inhibition of human recombinant renin assessed as decrease in plasma renin activity \\ by competitive radioimmunoassay in presence of human plasma\end{tabular} \vspace{0.2cm} \\ 
\hspace{0.5cm} B \hspace{0.5cm}     & CHEMBL1039373 & Inhibition of trypsin-activated human recombinant renin \vspace{0.2cm} \\ 
\hspace{0.5cm} C \hspace{0.5cm}     & CHEMBL2073500 & Inhibition of human KDR autophosphorylation expressed in mouse NIH/3T3 cells \vspace{0.2cm} \\ 
\hspace{0.5cm} D \hspace{0.5cm}     & CHEMBL2073498 & Inhibition of KDR by HTRF analysis in presence of 1 mM ATP \vspace{0.2cm} \\ 
\hspace{0.5cm} E \hspace{0.5cm}     & CHEMBL3804128 & \begin{tabular}[c]{@{}l@{}}Inhibition of wild type B-Raf in human MIAPaCa2 cells assessed as reduction in ERK \\ phosphorylation preincubated for 1 hr by Western blot method\end{tabular} \vspace{0.2cm} \\ 
\hspace{0.5cm} F \hspace{0.5cm}     & CHEMBL3804127 & \begin{tabular}[c]{@{}l@{}}Inhibition of B-Raf V600E mutant (unknown origin) assessed as MEK1 phosphorylation \\ using MEK1-Avitag as substrate after 1 hr by HTRF assay\end{tabular} \vspace{0.2cm}
\\ 
\bottomrule
\end{tabularx}
\caption{ChEMBL ID and assay description of assays A-F used to illustrate the different heterogeneity classes associated with the pairs (A,B) -- constant offset; (C,D) -- uncorrelated; (E,F) -- offset and rescaling.}
\label{si:assaydescription}
\end{table*}

\end{bibunit}